\def\gore{M.\,TOMCZAK: Flares that became XPEs}

\documentclass{ceab}   

\usepackage{epsfig}     
\usepackage{graphicx}   

\usepackage{ceabbib}     
\usepackage[T1]{fontenc}

\begin{document}

\title{Two solar flares that became X-ray plasma ejections}

\author{\normalsize  M.\,TOMCZAK \vspace{2mm} \\
        \it Astronomical Institute, University of Wroc{\l }aw, \\
        \it  ul.\,Kopernika 11, PL-51-622 Wroc{\l }aw, Poland}

\maketitle

\begin{abstract}
Solar flares and X-ray plasma ejections (XPEs) occur simultaneously
but usually are separated spatially. We present two exceptional
events observed by {\sl Yohkoh} in 2001 October 2 (event 1) and 2000
October 16 (event 2), in which features of flares and XPEs are
mixed. Namely, the soft and hard X-ray images show intense sources
of emission that move dynamically. Both events occurred inside broad
active regions showing complicated multi-level structure reaching up
to 200 Mm high. Both events show also similar four-stages evolution:
(1) a fast rise of a system of loops, (2) sudden changes in their
emission distribution, (3) a reconfiguration leading to liberation
of large amounts of plasma, (4) a small, static loop as the final
remnant. Nevertheless, the events are probably caused by different
physical processes: emerging magnetic flux plus reconnection (event
1) and reconnection plus ballooning instability (event 2). Different
is also the final destination of the ejected plasma: in the event 1
overlying magnetic fields stop the ejection, in the event 2 the
ejection destabilizes the overall magnetic structure and forms a
coronal mass ejection (CME).
\end{abstract}


\section{Introduction}

The solar corona is a place of restless interplay between plasma
flows and magnetic fields continuously modified by new flux emerging
from subphotospheric layers. X-ray radiation, emitted by hot plasma,
provides insight into localized episodes of magnetic energy
conversion. Sudden and temporal increases of brightness we call
flares, macroscopic motions of hot plasma we call X-ray plasma
ejections (XPEs).

Flares are more frequent than XPEs -- only 50-70\% of them are
associated with XPEs (Ohyama \& Shibata, 2000; Kim {\it et al.},
2005). On the other hand, if an XPE occurs, it does during the
impulsive phase of its associated flare (Tomczak \& Chmielewska,
2012). Plenty of movies collected in the XPE
catalog\footnote{http://www.astro.uni.wroc.pl/XPE/catalogue.html}
show that pairs of associated events (flare + XPE) are usually
separated spatially. It is in agreement with the {\sl canonical} 2D
CSHKP flare model ($\check{\rm{S}}$vestka \& Cliver, 1992 and
references therein), in which flare loops and a plasmoid occur on
opposite sides of the reconnection point. In a 3D model, an
expanding fluxtube has been proposed instead of the plasmoid
(Shibata {\it et al.}, 1995).

Flares and XPEs are usually easy to distinguish. XPEs are only
rarely observed as sources of hard X-ray emission. Also in soft
X-rays, flares are even several orders of magnitude brighter than
XPEs. An appropriate simultaneous imaging of both events often needs
different accumulation times of telescopes. On the other hand,
motion of XPEs with velocities of $10^2-10^3$~km\,s$^{-1}$ is easy
for detection in soft X-ray movies, in comparison with slowly rising
post-flare loop systems (several km\,s$^{-1}$).

In the XPE catalog we found two spectacular events, in which
features of flares and XPEs are mixed. They occurred on 2001 October
2 (event 1, the catalog number 346) and 2000 October 16 (event 2,
the catalog number 252). For both events we observe fast moving
magnetic structures, being bright sources of soft and hard X-ray
emission. This appearance seems to be in contradiction with the
standard flare model. Therefore our motivation is to propose a
consistent mechanism, explaining the observed features.

\section{INSTRUMENTS AND AVAILABLE DATA}

In our analysis we used observations derived by the following
instruments: (1) {\sl Yohkoh} Soft X-ray Telescope, SXT (Tsuneta
{\it et al.}, 1991), (2) {\sl Yohkoh} Hard X-ray Telescope, HXT
(Kosugi {\it et al.}, 1991), (3) {\sl Yohkoh} Bragg Crystal
Spectrometer, BCS (Culhane {\it et al.}, 1991), (4) {\sl SOHO}
Extreme ultraviolet Imaging Telescope, EIT (Delaboudiniere {\it et
al.}, 1995), (5) {\sl SOHO} Large Angle Spectroscopic Coronagraph,
LASCO (Brueckner {\it et al.}, 1995).

{\sl Yohkoh} provided the almost complete set of X-ray observations
illustrating the whole evolution of both analyzed events. SXT images
were made sequentially with three different spatial resolutions:
full resolution, FN, -- 2.45 arcsec, half resolution, HN, -- 4.9
arcsec, and quarter resolution, QN, -- 9.8 arcsec. A particular
resolution means a specific field of view: 2.6 $\times$ 2.6
arcmin$^2$, 5.2 $\times$ 5.2 arcmin$^2$, 10.5 $\times$ 10.5
arcmin$^2$, for the FN, HN, and QN resolutions, respectively. For
the FN resolution three different filters were used and the time
exposure was automatically adjusted. For the HN and QN resolutions
the time exposure was constant and long enough to record
weak-emission structures. In consequence, the brighter pixels were
heavily saturated. For the HN resolution two filters were used and
for the QN resolution -- only one.

Both events emitted enough photons in the energy bands L (14-23 keV)
and M1 (23-33 keV) of the HXT to reconstruct a sequence of hard
X-ray images. The number of counts in the band M2 (33-53 keV) is
limited -- only a few images can be made. There are no counts above
the background in the band H (53-93 keV).

Both events had also enough counts for allowing us to investigate
evolution changes of He-like ions recorded by the BCS. On the other
hand, the number of counts is luckily not so high to be suffered by
some instrumental effects like the line narrowing or the crystal
fluorescence (Trow {\it et al.}, 1994).

Important supplementary data were provided by the EIT and the LASCO
instruments, especially at large spatial scales. EIT imaged the
full-Sun images in ${\lambda}195$\AA\ with a 12-minutes cadence. The
LASCO coronagraphs C2 and C3 recorded on average 3 and 2 images per
hours, respectively.

Unfortunately, the {\sl TRACE} satellite provided its
high-resolution and high-cadence images of other active regions than
those in which the two events under study occurred.

\section{Event 1 (2001 OCTOBER 2)}

Event~1 occurred in the NOAA AR 9628 active region close to the
western solar limb. The estimated solar coordinates were
W86.4$\pm$2.9 S17.4 (Tomczak, 2009). The light curves recorded by
the {\sl Geostationary Operational Environmental Satellites} ({\sl
GOES}) show a fast rise to the short-term maximum at 17:16 UT ({\sl
GOES} class C4.7) followed by a slow decay lasting about four hours.

\begin{figure}[t]
\epsfig{file=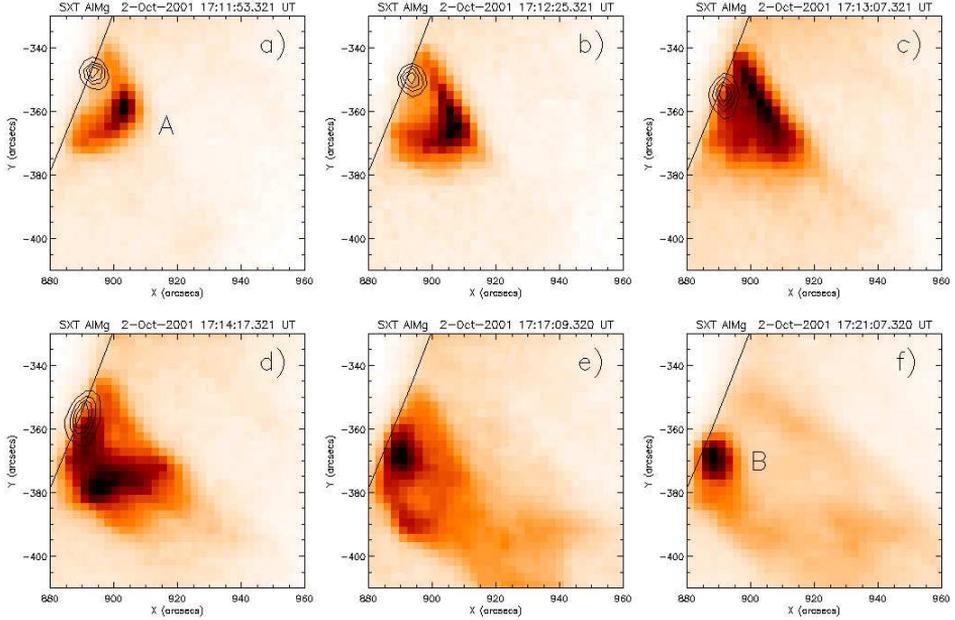,width=12.5cm} \caption{Early
development of the event of 2001 October 2. Soft X-ray emission
(SXT/AlMg filter, 2.4-32 \AA\ , images) is illustrated by reverse
halftones, hard X-ray emission (HXT/M1 energy band, 23-33 keV,
images) is overplotted as isocontours. The solar limb is plotted by
a straight black line.}
\end{figure}

The six SXT snapshots illustrating the early evolution of the event
are given in Figure\,1\footnote{Full movies are available in the XPE
Catalog: event number 346, entry XPE}. The following four stages of
evolution can be distinguished:
\begin{enumerate}
\item A fast rise of the loop A (17:11.5-17:13 UT, Figures 1a-b):
\newline
This loop is responsible for the maximum seen in the {\sl GOES}
light curves. The progressive increase of the apex height is
associated with the progressive decrease of the distance between the
loop footpoints. The velocity of the apex height rises from 75 to
165 km\,s$^{-1}$, the velocity of the footpoint approach decreases
from 60 to 20 km\,s$^{-1}$. The above changes cause an increase of
the loop's curvature, described by the ratio between height and
diameter. The ratio changed from 0.3 to 1.0. The loop had a circular
shape ($H/D = 0.5$) around 17:12 UT. At the same moment the loop was
almost symmetrical, as suggests the ratio $D_1/D_2 \sim 1$, where
$D_1$ and $D_2$ are the distances between projection of the loop
apex on the solar surface and the north or south footpoint,
respectively. Earlier and later the apex was shifted to the south
reaching the maximum value ${\sim}2$ at 17:13 UT just before the
reconfiguration.
\item Sudden changes in the emission distribution inside the loop A
(17:12.5-17:13 UT, Figure 1c): \newline The typical bright loop-top
vanishes and the almost uniformly distributed emission concentrates
in the northern leg.
\item The main reconfiguration (17:13-17:16 UT, Figures 1d-e):
\newline The loop A seems not to exist anymore. Two large radial
structures become seen in distance about 15 Mm one from to other. A
plasma outflow is clearly seen along them. The previous loop A was
contained somewhere between these structures.
\item The long-lived remnant (after 17:16 UT, Figure 1f):
\newline A small loop B at the place of the southern footpoint of
the loop A is the brightest. The loop does not change distinctly its
size and shape.
\end{enumerate}

The filter ratio method allows us to estimate amounts of plasma
inside the loop A and the loop B. We obtain $4 \times 10^{13}$ and
$2 \times 10^{13}$ grams, respectively. It means that at least $2
\times 10^{13}$ grams of plasma was liberated due to the reported
reconfiguration.

The HXT images reconstructed in the energy bands L, M1, and M2
accordingly localize the main hard X-ray source at the solar limb,
somewhere between the footpoints of the loop A (see isolines in
Figure\,1). The hard X-ray spectra are relatively flat at the
beginning (17:11:20 UT) with the power-law indices $\gamma$ of about
4. The progressive steepening leads to values of $\gamma$ of about 8
at 17:16 UT.

The high-resolution spectra of He-like ions of Fe, Ca, and S
recorded by the BCS show that the observed lines are extremely
broadened within time interval between 17:11.5 and 17:14.5 UT. The
calculated values of the non-thermal turbulent plasma motions reach
even 600 km\,s$^{-1}$ for Fe~{\sc xxv} and Ca~{\sc xix} ions, and
400 km\,s$^{-1}$ for S~{\sc xv}. For flares these values are usually
2-3 times lower (e.g., Alexander {\it et al.}, 1998; Mariska \&
McTiernan, 1999).

Further evolution of the plasma liberated from the loop A, that did
not join the loop B, can be observed in the SXT images of the HN and
QN resolution having broader field of view. First, the response of
many loops of heights below 90 Mm is seen. After that the systematic
expansion along a narrow, radial structure between heights 100 and
190 Mm occurs, for the time interval 17:24-17:37 UT. It gives a
modest average velocity of about 110 km\,s$^{-1}$. Further expansion
fades above the level 200-250 Mm.

\begin{figure}[t]
\epsfig{file=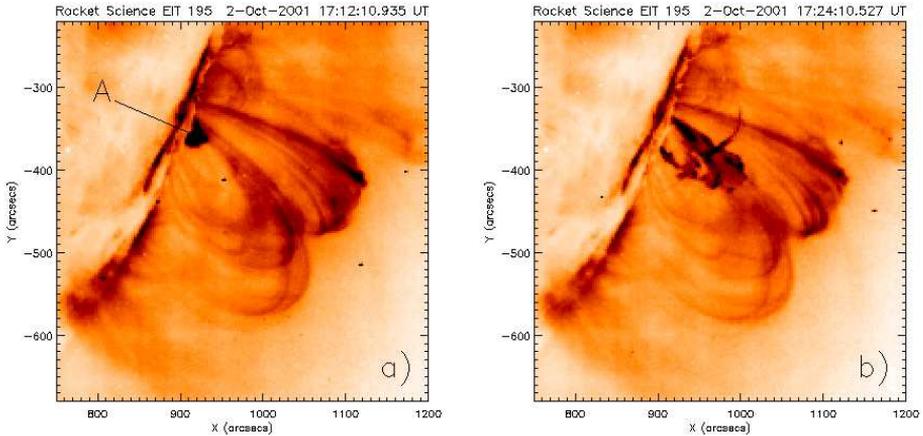,width=12.5cm} \caption{Image recorded
by EIT in the 195 \AA\ passband during the early development of the
event of 2001 October 2. Brightness scale is reversed. The loop A is
marked with an arrow.}
\end{figure}

The images acquired by EIT allow us to better understand the overall
magnetic structure of the active region. Figure~2 shows two of them
taken during the early evolution of the event. It is seen that the
loop A is localized close to the footpoints of the high loops
forming an arcade reaching altitude of 170 Mm. Next snapshots, not
included in this paper, show that the low-altitude reconfiguration
does not modify the arcade with the exception of some oscillations.
Unfortunately, the sparse, 12-minutes cadence does not allow to
calculate the period of oscillations.

No coronal mass ejection (CME) nor a radio bursts were connected
with the described event.

\section{Event 2 (2000 October 16)}

This event occurred in the NOAA AR 9182 active region behind the
western solar limb. The estimated solar coordinates were
W107.4$\pm$3.3\,N01.9 (Tomczak, 2009). It means that only the higher
parts of this active region (above 34 Mm) could be seen. Light
curves recorded by ({\sl GOES}) show a fast rise between 05:33 and
05:40 UT, a flat maximum around 05:49 UT ({\sl GOES} class C7.0),
then a slower decay until 06:35 UT. Thereafter, another flare
occurred in the same active region. It was stronger (M2.5 class) and
lasted about 9 hours, therefore the described event can be
considered as its precursor.

\begin{figure}[t]
\epsfig{file=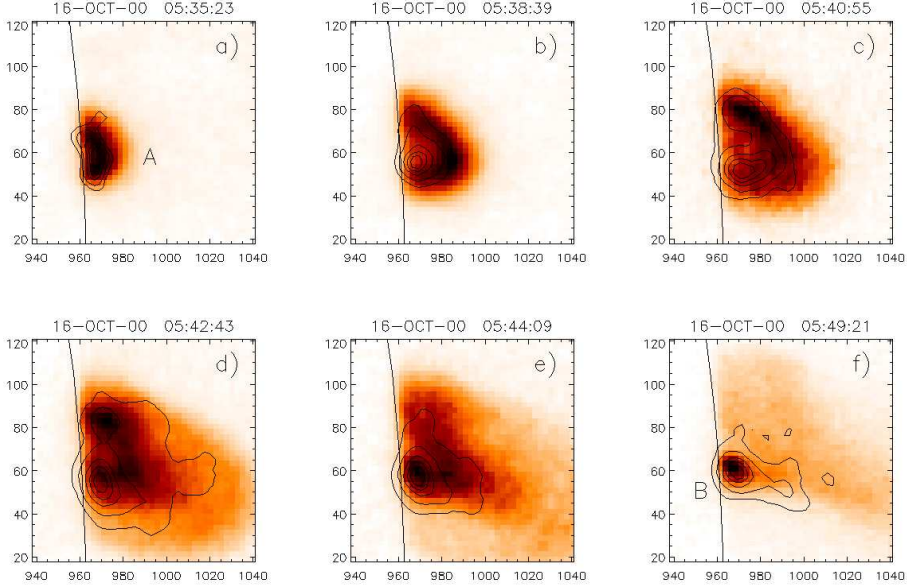,width=12.5cm} \caption{Early
development of the event of 2000 October 16. Soft X-ray emission
(SXT/Be119 filter, 2.3-10 \AA\ , images) is illustrated by reverse
halftones, hard X-ray emission (HXT/L energy band, 14-23 keV,
images) is overplotted as isocontours. The solar limb is plotted by
a straight black line.}
\end{figure}

The six SXT snapshots illustrating the early development of the
event are given in Figure\,3\footnote{Full movies are available in
the XPE Catalog: event number 252, entry XPE}. Similar four-stages
evolution, like for the event 1, can be distinguished:
\begin{enumerate}
\item A fast rise of the loop A (05:33-05:39.5 UT, Figures 3a-b):
\newline This loop is responsible for the maximum seen in the {\sl GOES} light
curves. The apex rises almost constantly and its velocity does not
exceed 100 km\,s$^{-1}$. The curvature of the apex stays
approximately the same.
\item Sudden changes in the emission distribution inside the loop A
(05:39.5-05:41 UT, Figure 3c):
\newline Instead of the bright loop-top the northern leg becomes most dominant.
The movies make an impression of a massive plasma downflow.
\item The main reconfiguration (05:41-05:46 UT, Figures 3d-e):
\newline The velocity of the apex rise suddenly increases above 200 km\,s$^{-1}$,
then a modest deceleration ($\sim0.3$ km\,s$^{-2}$) occurs. A strong
emission drop of fast expanding apex together with a gradual
concentration of the bright emission close above the solar limb make
an impression of plasma liberation from the loop A. However, in
spite of the progressive growth the initial shape of the loop is
still recognizable.
\item The long-lived remnant (after 05:46 UT, Figure 3f):
\newline A small loop B inside the inner contour of
the loop A at the beginning of the stage 1 is the brightest. The
loop B does not change distinctly its size and shape.
\end{enumerate}

The filter ratio method allows us to estimate amounts of plasma
inside the loop A and the loop B. We obtain $2 \times 10^{14}$ and
$1 \times 10^{13}$ grams, respectively. It means that at least $1.9
\times 10^{14}$ grams of plasma was liberated due to the reported
reconfiguration.

The HXT images reconstructed in the energy bands: L, M1, and M2 show
that the loops A and B are also main hard X-ray emission sources
(see isolines in Figure 3). The hard X-ray spectra are relatively
soft (power-law indices $\gamma$ are between 7 and 9), which can be
considered as typical appearance of coronal sources (Tomczak, 2009).
The three weak increases of hard X-ray flux observed in the energy
band M2 occurred at 05:38, 05:43, and 05:48 UT.

The spectra of He-like ions Fe, Ca, and S recorded by the BCS show
that the observed lines are very broad within the time interval
between 05:36 and 05:46 UT. The calculated values of non-thermal
turbulent plasma motions reach even 400, 250, and 170 km\,s$^{-1}$
for Fe~{\sc xxv}, Ca~{\sc xix}, and  S~{\sc xv}, respectively. The
time changes of the non-thermal motions have a complex look with
three maxima at 05:38, 05:41.5, and 05:44 UT.

Further evolution of the plasma liberated from the loop A, that did
not join the loop B, can be observed in the SXT images of the HN and
QN resolution having broader field of view. The systematic expansion
of a loop-like structure until 06:05 UT, when {\sl Yohkoh} came into
the Earth's shadow, is seen. The final altitude is above 200 Mm,
which gives an average velocity below 100 km\,s$^{-1}$. The
expansion occurs south to the arcade of loops of the heights between
80 and 140 Mm.

\begin{figure}[t]
\epsfig{file=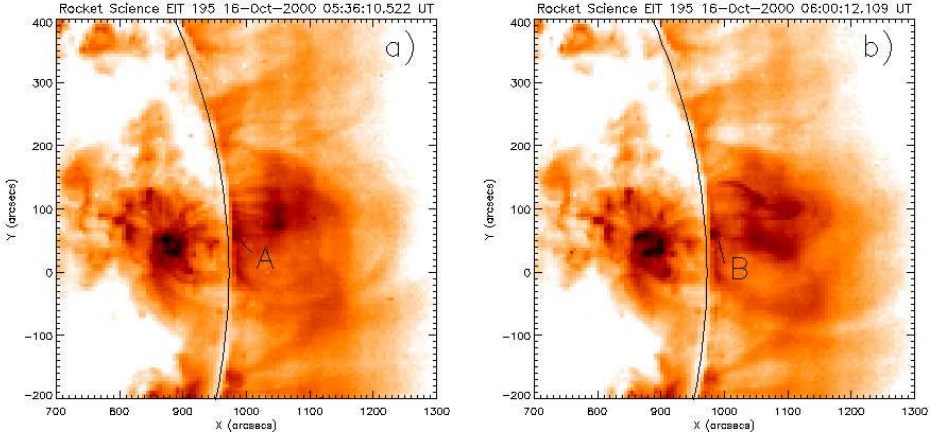,width=12.5cm} \caption{Images recorded
by EIT in the 195 \AA\ passband during the development of the event
of 2000 October 16. Brightness scale is reversed. The loops A and B
from Figure 3 are marked.}
\end{figure}

The EIT 195 \AA\ images allow us to better understand the overall
magnetic structure of the active region. Figure~4 shows two of them
taken during the development of the event. It is seen that the
expansion caused by the reconfiguration of the loop A destabilizes
the arcade already existing to the north. Next snapshots, not
included in this paper, show the violent explosion that occurred
about 06:36 UT. It is followed by the CME, reported in the SOHO
LASCO CME Catalog\footnote{http://cdaw.gsfc.nasa.gov/CME\_list/}
(Gopalswamy {\it et al.}, 2009), as a fast ($\sim1400$
km\,s$^{-1}$), halo event.

\section{Discussion}

At the first look, the described events seem to be quite similar,
nevertheless, in our opinion different mechanisms are responsible
for their occurrence. Therefore, they are discussed in separate
subsections.

\subsection{Event 1}

The magnetic configuration, in which the event of 2001 October 2
occurred, resembles that introduced in the Emerging Flux Model
(Heyvaerts {\it et al.}, 1977). In this model, subphotospheric
magnetic fields emerge due to the buoyancy effect and meet overlying
coronal magnetic fields. A current sheet is formed and reconnection
process is responsible for energy release and acceleration of
electrons.

In the described event the loop A and the legs of some loops seen in
the EIT image represent emerging and overlying coronal fields,
respectively. The reconnection develops since 17:11 UT and the loop
A is filled with hot, dense plasma due to chromospheric evaporation,
which makes the loop bright in soft X-rays. The accelerated
electrons are stopped at the entrance to the chromosphere producing
hard X-ray emission. When the inflow of subphotospheric fields
weakens, the loop A begins to deform. The further reconnection
destroys the loop and either the liberated plasma falls downward to
the loop B, that is a postflare loop, or expands outward along
magnetic lines of the overlying arcade. These lines protect against
the final evacuation of plasma from the Sun (no CME occurs). The
flows are generally sub-Alfv$\acute{\rm{e}}$nic, therefore no radio
type II burst was reported.

\subsection{Event 2}

We consider the loop A, seen at the beginning of the event of 2000
October 16, as an unusual example of an XPE formed due to the
magnetic reconnection behind the solar limb. XPEs are rather faint
sources of X-ray emission, therefore the loop A resembles a flare.
Further expansion of the loop A leads probably to an instability.

Among from macroscopic hydrodynamic and MHD plasma instabilities
relevant in coronal loops (Aschwanden, 2004), the ballooning
instability seems to be the most promising. The ballooning
instability is a kind of interchange instabilities occurring at the
interface between two fluid layers. For this particular instability
the interface between two layers, where the plasma beta is low
($\beta < 1$) and high ($\beta > 1$), is important. The ballooning
instability is often recognized in tokamak fusion power reactors and
in the Earth's magnetosphere but rather exceptionally in the solar
corona.

Shibasaki (2001) discussed the interplay of outward (pressure
gradient, magnetic pressure gradient, centrifugal force) and inward
(magnetic tension, gravitation) forces in a context of the
equilibrium of plasma in a coronal loop. We used formulae developed
by Shibasaki to quantify the most important parameters responsible
for the balance in the described event, i.e. centrifugal force of
thermal plasma flows along the loop curvature versus magnetic
tension force.

Crossing of a terminal value of the plasma beta triggers in the
ballooning instability further evolution, therefore we assumed that
$\beta = 1$ at the apex of the loop A just before the main
reconfiguration. In this way it is possible to estimate the magnetic
field strength. For the temperature, $T = 13.4$ MK, and the electron
number density, $N_e = 2.6 \times 10^{10}$ cm$^{-3}$, calculated
according to the filter ratio method, we obtain $B = 50\pm5$ G. We
adopted the dipole field formula, $B(h) \approx B_0(1 +
h/h_D)^{-3}$, where $h_D = 75$ Mm, proposed by Aschwanden {\it et
al.} (1999), to calculate plasma beta maps during the main
reconfiguration. The results are consistent with the ballooning
instability interpretation -- there is a permanent area of
high-$\beta$ around the apex of the expanding loop-like structure,
of which the loop A was a progenitor.

The detailed numbers show that the increase of the plasma beta was
mainly caused by a drop of the magnetic field strength due to the
expansion. However, the symptoms of energy release like the electron
number density increase and the presence of turbulent motions should
be also taken into consideration.

\section{Conclusions}

The two described events show that the division between XPEs and
flares can be artificial, to some extent. Magnetic reconnection
associated with hydrodynamic response on energy release can magnetic
loops bring either a loss-of-equilibrium or an inflow of dense,
chromospheric plasma. In the first case we observe an XPE, whereas
in the second case -- a flare. Usually XPEs and flares are spatially
separated because magnetic structures above the reconnection point
are weaker than those situated below. The described events are
unique because magnetic loops strong enough to keep hot, dense
plasma became weaker due to the interaction with an overlying
coronal field (event 1) or due to expansion (event 2).

The overall structure of active regions plays very important role in
the final destination of plasma ejected due to instabilities in the
lower corona. Even similar velocities of expansion lead to opposite
scenarios: full confinement (event 1) versus CME formation (event
2).

\section*{Acknowledgements}
{\sl Yohkoh} is a project of the Institute of Space and
Astronautical Science of Japan. {\sl SOHO} is a project of
international cooperation between ESA and NASA. We acknowledge
useful comments made by the Referee and financial support from the
Polish National Science Centre grant 2011/03/B/ST9/00104.


\section*{References}
\begin{itemize}
\small
\itemsep -2pt
\itemindent -20pt
\item[] Alexander, D., Harra-Murnion, L.\,K., Khan, J.\,I.,
Matthews, S.\,A.: 1998, {\it Astrophys.\,J.\,Lett.}, {\bf 494},
L235.
\item[] Aschwanden, M.\,J.: 2004, {\it Physics of the Solar Corona.
An Introduction}, Springer: Praxis.
\item[] Aschwanden, M.\,J.,  Newmark, J.\,S., Delaboudiniere, J.-P., {\it et al.}:
1999, {\it Astrophys.\,J.}, {\bf 515}, 842.
\item[] Brueckner, G.\,E., {\it et al.}: 1995, {\it Solar Phys.}, {\bf 162},
357.
\item[] Culhane, J.\,L., {\it et al.}: 1991, {\it Solar Phys.}, {\bf
136}, 89.
\item[] Delaboudiniere, J.-P., {\it et al.}: 1995, {\it Solar Phys.},
{\bf 162}, 291.
\item[] Gopalswamy, N., {\it et al.}: 2009, {\it Earth Moon Planet}, {\bf 104},
295.
\item[] Heyvaerts, J., Priest, E.\,R., Rust, D.\,M.: 1977, {\it Astrophys.\,J.},
{\bf 216}, 123.
\item[] Kim,\,Y.-H., Moon,\,Y.-J., Cho,\,K.-S., Kim, K.-S., Park, Y.\,D.:
2005, {\it Astrophys.\,J.}, {\bf 622}, 1240.
\item[] Kosugi, T., {\it et al.}: 1991, {\it Solar Phys.}, {\bf
136}, 17.
\item[] Mariska, J.\,T., McTiernan, J.\,M.: 1999, {\it
Astrophys.\,J.}, {\bf 514}, 484.
\item[] Ohyama, M., Shibata, K.: 2000, J.\,Atm. Solar Terr.\,Phys., {\bf 62},
1509.
\item[] Shibasaki, K.: 2001, {\it Astrophys.\,J.}, {\bf 557}, 326.
\item[] Shibata,\,K., Masuda,\,S., Shimojo,\,M., {\it et al.}: 1995,
{\it Astrophys.\,J.\,Lett.}, {\bf 451}, L83.
\item[] $\check{\rm{S}}$vestka, Z., Cliver, E.\,W.: 1992, in IAU Coll. 133,
{\it Eruptive Solar Flares}, ed. Z. $\check{\rm{S}}$vestka, B.\,V.
Jackson, \& M.\,E. Machado (New York: Springer), 1.
\item[] Tomczak, M.: 2009, {\it Astron. Astrophys.}, {\bf 502}, 665.
\item[] Tomczak, M., Chmielewska, E..: 2012, {\it
Astrophys.\,J.\,Suppl.}, {\bf 199}, 10.
\item[] Trow, M.\,W., Bento, A.\,C., Smith, A.: 1994, {\it Nucl.
Instrum. Methods Phys. Res.}, {\bf 348}, 232.
\item[] Tsuneta,\,S., {\it et al.}: 1991, {\it Solar Phys.},
{\bf 136}, 37.
\end{itemize}

\bibliographystyle{ceab}
\bibliography{sample}

\end{document}